# A large magneto-inductance effect in $La_{0.67}Ba_{0.33}MnO_3$

V. B. Naik, A. Rebello and R. Mahendiran[1]

Department of Physics and NUS Nanoscience and Nanotechnology Initiative (NUSNNI), Faculty of Science, National University of Singapore, 2 Science Drive 3, Singapore 117542, Singapore

**Abstract**

We report four probe impedance ($Z = R+iX$) of $La_{0.67}Ba_{0.33}MnO_3$ at $f$ = 100 kHz under different dc bias magnetic fields. The ac resistance (R) exhibits a peak around $T_P$ = 325 K which is accompanied by a rapid increase and a peak in the reactance (X) in a zero field. The magnetoreactance ($\Delta X/X$) exhibits a sharp peak close to $T_P$ and its magnitude (≈ 60% in $H$ = 1 kG) exceeds that of the *ac* magnetoresistance ($\Delta R/R$ = 5 %). It is suggested that the magnetoreactance arises from changes in the self inductance of the sample rather than the capacitance.

---

[1] E-mail: phyrm@nus.edu.sg

In recent years, multiferroic materials which show coupled magnetic and electric dipoles ordering have received a great attention due to the possibility of controlling the ferroelectric polarization with magnetic field and magnetization with electric field.[1] One of the indications of the coupled magnetic and electrical ordering is the appearance of an anomaly in the dielectric constant ($\varepsilon$) at the magnetic phase transition temperature while varying the temperature.[2] The $\varepsilon$ can increase or decrease under an external magnetic field. A large magnetodielectric response around the magnetic phase transition has been reported in manganites of the formula $RMnO_3$ where R = Tb, Gd, Dy etc.,[3] and perovskites containing Mn and Ni or Fe ions.[4] These oxides are highly insulating and show negligible magnetoresistance. However, a large increase in the dielectric constant ($\Delta\varepsilon/\varepsilon$ = 60 % at $\mu_0H$ = 0.5 T) was also reported in $La_{0.67}Ca_{0.33}MnO_3$[5] just above the ferromagnetic Curie temperature ($T_C$) and in the phase separated $Pr_{0.67}Ca_{0.33}MnO_3$[6] at low temperature. G. Catalan[7] has suggested that the magnetocapacitance in these materials could be an artifact of magnetoresistance and can be understood through the Maxwell-Wagner relaxation model developed for inhomogeneous semiconductors. A recent theoretical prediction suggests possible occurrence of the magnetocapacitance effect even in the non magnetic composite media.[8]

In dielectric measurements, one generally measures the real (Z') and imaginary (Z") components of the total impedance (Z = Z'+iZ"). From the measured impedance, the real ($\varepsilon'$) and imaginary ($\varepsilon''$) components of the dielectric permittivity are calculated using the relations $\varepsilon' = Z''/(\omega C_0 Z^2)$ and $\varepsilon'' = Z'/(\omega C_0 Z^2)$ where $C_o$ is the capacitance of the empty cell with the dimensions of the sample.[9] However, if a sample is magnetic and has low resistivity, the measured Z' and Z" can have contribution from the magnetic permeability which may exceed that of the dielectric permittivity. In this case, the changes in Z' and Z" under an external magnetic field are not necessarily due to the magnetocapacitance effect, but can be caused by variation of magnetic permeability. Although the idea is simple, it has not yet been investigated carefully. The purpose of this paper is to show that besides the magnetocapacitance effect, another phenomenon called magnetoinductance effect occurs in the conducting ferromagnet $La_{0.67}Ba_{0.33}MnO_3$.

We have measured the alternating current (ac) impedance ($Z = R+iX$) of a polycrystalline $La_{0.67}Ba_{0.33}MnO_3$ of rectangular bar shaped sample (7x3.3x2 mm$^3$) using an Agilent LCR meter (model 4285A) with 20 mV excitation at a constant frequency ($f =$ 100 kHz) in linear four probe configuration. The Physical Property Measurement System (PPMS, Quantum Design Inc) equipped with commercial superconducting magnet was used for varying temperature ($T = 350$ K – 10 K) and magnetic fields ($0 < H < 1$ kG). The magnetic field was applied along the direction of the current through the sample. The ac susceptibility ($f = 10$ kHz and ac magnetic field is 0.2 G) at different longitudinal dc bias magnetic fields was studied by measuring the self-inductance ($L_{ac}$) of a 50 turns single layered solenoid loaded with the sample using the National Instrument PXI 4072 LCR meter plug-in card.

Figure 1(a) shows the temperature dependence of the *dc* resistivity of $La_{0.67}Ba_{0.33}MnO_3$ in zero magnetic field and $\mu_0H = 7$ T (left scale). The sample undergoes a paramagnetic semiconductor to a ferromagnetic metal transition in a zero magnetic field with a peak in $\rho(T)$ at $T_P$ = 325 K. The dc magnetoresistance at $\mu_0 H$ = 7 T shown on the right scale exhibits a peak ($\Delta R/R$ = 47 %) close to 325 K and it is 50 % at $T$ = 10 K. It is known that the dc magnetoresistance for $T \ll T_C$ in ferromagnetic manganites is an extrinsic effect which arises from spin-polarized tunneling of charge carrier between the ferromagnetic grains through spin disordered grain boundaries.[10] Figure 1(b) shows the temperature dependence of the ac inductance of the sample measured at $f$ = 10 kHz using a solenoid under different dc bias fields applied parallel to the long axis of the coil. The ac inductance is small and is independent of temperature above 330 K, but it increases rapidly and exhibits a peak very near to $T_P$ in the absence of an external magnetic field. The $T_C$ determined from the inflection point of the rapidly increasing portion of the inductance curve is 321 K which is 4 K below the peak in the resistivity ($T_P$). The peak decreases in magnitude, broadens and shifts down in temperature with increasing strength of the magnetic field. The suppression and broadening of the susceptibility peak is due to the inability of the small ac magnetic field to induce domain wall motion when the direction of domain magnetization is fixed by the larger dc magnetic field. As the dc magnetic field strength increases, domain rotation and the effect of ac magnetic field in reversing the domain magnetization have also to be considered. Note that the longitudinal susceptibility is primarily affected by the dc magnetic field in contrast to the transverse susceptibility when the current directly passes through the sample.

Figure 2 (a) shows the ac resistance (R) at $f$ = 100 kHz under different dc magnetic fields applied parallel to the direction of the ac current in the sample. The R in a zero field exhibits a peak at $T_P$ = 325 K. While the peak is hardly affected as $H$ increases from 0 G to 1 kG, the *ac* resistance below 300 K decreases slightly. However, the peak decreases in magnitude and shifts to higher temperature as *H* increases beyond 1 T as like in the *dc* resistivity. Figure 2 (b) shows temperature dependence of the reactance (X). The X in a zero field is nearly independent of temperature above 330 K, but it steeply increases around $T_C$ = 321 K and shows a peak close to it. The $T_C$ is estimated from the inflection point of the X-T curve. Below the peak, X decreases down to 150 K and then shows a tendency to increase with further decrease in *T*. The X is one order of magnitude lower than R. Very interestingly, the X, unlike R, is very sensitive to sub-kilo gauss magnetic field. The peak decreases in magnitude and broadens as *H* increases from 0 to 500 G. The peak is no more visible for *H* = 700 G or above. Instead, we notice only a gradual increase of X with lowering temperature below 325 K. A similar behavior is seen in X up to $\mu_0 H$ = 7 T.

Figure 3(a) shows the percentage change in the ac magnetoresistance ($\Delta R/R$ = [R(*H*)-R(0)]/R(0)) and figure 3(b) shows the magnetoreactance ($\Delta X/X$ = [X(*H*)-X(0)]/X(0)) under different *dc* magnetic fields for $H \leq 1$ kG. The value of $\Delta R/R$ at *H* = 100 G is very small and does not show a peak around $T_C$. As *H* increases, a peak in $\Delta R/R$ emerges at $T_C$. In addition, the ac magnetoresistance at the lowest temperature also increases in magnitude. For instance, the peak at $T_C$ attains a maximum value of 5 % at 1 kG, but it increases to ≈ 22 % at *T* = 10 K. On the other hand, $\Delta X/X$ shows a sharp peak around $T_C$ and decreases with lowering temperature. The magnitude of the peak in $\Delta X/X$ increases from ≈ 10 % at *H* = 100 G to a huge value of ≈ 60 % at *H* = 1 kG. Thus, the magnitude of the magnetoreactance far exceeds that of the ac magnetoresistance at the same field strength. The position of the ΔX/X peak is not much affected by the magnetic field. We find that the value of ΔX/X at the lowest temperature also increases in magnitude as $H$ increases from 100 G to 1 kG, but it is still lower than its peak value at all values of $H$ below 1 kG. Fig. 3(c) shows the magnetoimpedance (ΔZ/Z) which reaches a maximum value of 5 % at the peak at $H$ = 1 kG and it is dominated by the behavior of the *ac* magnetoresistance. The observed ΔZ/Z at $f$ = 100 kHz and $T$ = 200 K is slightly lower than 10 % reported in a similar composition by Hu *et al.*,[11] but their samples showed a semiconductor-metal transition around 200 K which was much below the onset of ferromagnetic transition.

It is known that the low field magnetoresistance in polycrystalline manganites arises from tunneling of the spin polarized $e_g$-holes between the ferromagnetic grains separated by high resistive grain boundaries. The monotonic increase of ΔR/R below the peak suggests that the tunneling magnetoresistance is still active at $f$ = 100 kHz. We are more concerned with the huge magnetoreactance observed. Generally, reactance in polycrystalline manganites is attributed to intra- and inter-grain (grain boundary) capacitances. Glazer and Zieze[12] showed that the capacitance of grain boundaries formed at step-edges in $La_{0.7}Ca_{0.3}MnO_3$ film is negligible in the paramagnetic state but increases below the $T_C$. Their model indicated that the capacitance increases with the second power of the difference between grain and grain-boundary magnetization below $T_C$. They concluded that the capacitance follows more closely the grain boundary magnetization. Although it is tempting to attribute the observed magnetoreactance to magnetocapacitance effect, we would like to consider other possibility. It is known that if an *ac* current, $I(t) = I_0 e^{i\omega t}$ flows through a ferromagnetic metal of resistance R, the *ac* voltage drop across a ferromagnetic conductor is not simply $V(t) = I(t)R$ but $V(t) = I(t)R + L dI(t)/dt$, where L is the self inductance of the sample. Thus, the impedance is $Z = V(t)/I(t) = R + j\omega L = R + jX$. Hence, the reactance is proportional to the self inductance of the sample, which in turn proportional to the initial permeability of the sample. Similar results can be also obtained from the classical theory of electromagnetic wave propagation in solids.

As we know, the flow of radio frequency (*rf*) current in a ferromagnetic material is confined to a surface layer of thickness known as skin depth or penetration depth and is given by $\delta = \sqrt{(2\rho)/(\omega \mu_t)}$, where $\rho$ is the *dc* resistivity, $\omega$ is the angular frequency of the *rf* current and $\mu_t$ is the effective transverse initial permeability. We consider the transverse permeability instead of the longitudinal permeability, because the flow of *ac* current along the long axis of the sample induces an *ac* magnetic field in the transverse direction to the current. The non magnetic skin depth (assuming $\mu_t = 1$) at 300 K in our sample is 56.05 mm at 100 kHz, which is much larger than the thickness of the sample (2t = 3 mm). From the classical theory of electromagnetism, electrical impedance of a current carrying conductor of cylindrical geometry with radius 2a is obtained as $Z/R_{dc} = (ka)J_0(ka)/2J_1(ka)$, where $R_{dc}$ is the *dc* resistance, $k = (1-j)/\delta$, $J_0$ and $J_1$ are zero and first-order Bessel functions.[13] If the skin effect is weak, then the Bessel functions can be expanded to give $Z' = R_{dc}[1 + 1/48(a/\delta)^4] = R_{dc}[1 + (\pi\mu_0\mu_t\sigma\omega a^2)^2/12]$ and $Z'' = X = \tfrac{1}{4}R_{dc}(a/\delta)^2 \propto (\mu_t\omega)$, where $Z = Z' + jZ''$. At low frequencies such as $f = 100$ kHz which corresponds to the weak skin effect, the $(a/\delta)^4$ term in Z' is smaller than $R_{dc}$ and hence it has negligible effect on the *ac* resistance. However, $X \propto \mu_t$, where $X = \omega L = \omega G \mu_t$ (G is the geometrical factor) and X is the dominant factor. Hence the observed peak in X around $T_C$ and its suppression in a small magnetic field can be understood in terms of magnetoinductance. The similarity between the observed *ac* inductance using coil and X under different biased *dc* magnetic fields is in supportive of the magnetoinductance effect rather than the magnetocapacitance effect.

In summary, our investigation of ac electrical transport in $La_{0.67}Ba_{0.33}MnO_3$ at $f =$ 100 kHz indicates a large low field magnetoreactance ($\Delta X/X = 60$ % at $T_C$ and $H = 1$ kG) which exceeds that of the ac magnetoresistance ($\Delta R/R = 5$ %). The observed magnetoreactance is attributed to the change in the magnetic permeability ("magnetoinductance") of the sample rather than the variation in the dielectric permittivity ("magnetocapacitance"). Such a large low field magnetoreactance may find applications in thin film inductor. Further studies over a wide frequency range could be helpful to understand the competition between the magnetocapacitance effect found in phase separated manganites such as bulk Cr doped $Nd_{0.5}Ca_{0.5}MnO_3$[14] and La-Pr-CaMnO$_3$ thin film[15] and the magnetoinductance effect reported here.

Acknowledgements: R. M. acknowledges NUS and MOE (Singapore) for supporting this work through the grants R-144-000-197-123 and AFC/Tier-1-R-144-000-167-112.

**References:**

[1] R. Ramesh and N. A. Spaldin, Nat. Mater. **6**, 21 (2007) and references therein.

[2] T. Kimura, S. Kawamoto, I. Yamada, M. Azuma, M. Takano and Y. Tokura, Phys. Rev. B **67**, 180401 (2003); Y. Yang, J.-M. Liu, H. B. Huang, W. Q. Zou, P. Bao and Z. G. Liu, Phys. Rev. B **70**, 132101 (2004).

[3] N. Hur, S. Park, P. A. Sharma, J. S. Ahn, S. Guha and S.-W. Cheong, Nature (London) **429**, 392 (2004) and references therein.

[4] M. P. Singh, K. D. Truong and P. Fournier, Appl. Phys. Lett. **91**, 042504 (2007); N. S. Rogado, J. Li, A. W. Sleight and M. A. Subramanian, Adv. Mater. **17**, 2225 (2005); P. Padhan, H. Z. Guo, P. LeClair and A. Gupta, Appl. Phys. Lett. **92**, 022909 (2008); A. K. Kundu, R. Ranjith, B. Kundys, N. Nguyen, V Caignaért, V Pralong, W. Prellier and B. Raveau, Appl. Phys. Lett. **93**, 052906 (2008).

[5] J. Rivas, J. Mira, B. Rivas-Murias, A. Fondado, J. Dec, W. Kleemann and M. A. Señaris-Rodríguez, Appl. Phys. Lett. **88**, 242906 (2006).

[6] S. Mercone, A. Wahl. A. Pautrat, M. Pollet and C. Simon, Phys. Rev. B **69**, 174433 (2004); R. S. Freitas, J. F. Mitchell and P. Schiffer, Phys. Rev. B 72, 144429 (2005).

[7] G. Catalan, Appl. Phys. Lett. **88**, 102902 (2006).

[8] M. M. Parish and P. B. Littlewood, Phys. Rev. Lett. **101**, 166602 (2008).

[9] D. C. Sinclair and A. R. West J. Appl. Phys. **66**, 3850 (1989).

[10] H. Y. Hwang, S. W. Cheong, N. P. Ong and B. Batlogg, Phys. Rev. Lett. **77**, 2041 (1996).

[11] J. Hu, H. Qin, H. Niu, L. Zhu, J. Chen, W. Xiao and Y. Pei, J. Magn. Magn. Mater. **261**, 105 (2003).

[12] A. Glaser and M. Ziese, Phys. Rev. B **66**, 094422 (2002).

[13] L. V. Panina, K. Mohri, K. Bushida and M. Noda, J. Appl. Phys. **76,** 6198 (1994); K. R. Pirota, M. Knobel, and C. Gomez-Polo, Physica B, **320**, 127 (2002).

[14] A. S. Carneiro, F. C. Fonseca, R. F. Jardim and T. Kimura  J. Appl. Phys. **93**, 8074 (2003).

[15] R. P. Rairigh, G. S. Bhalla, S. Tongay, T. Dhakal, A. Biswas  and A. F. Hebard, Nature Physics, **3**, 551 (2007).

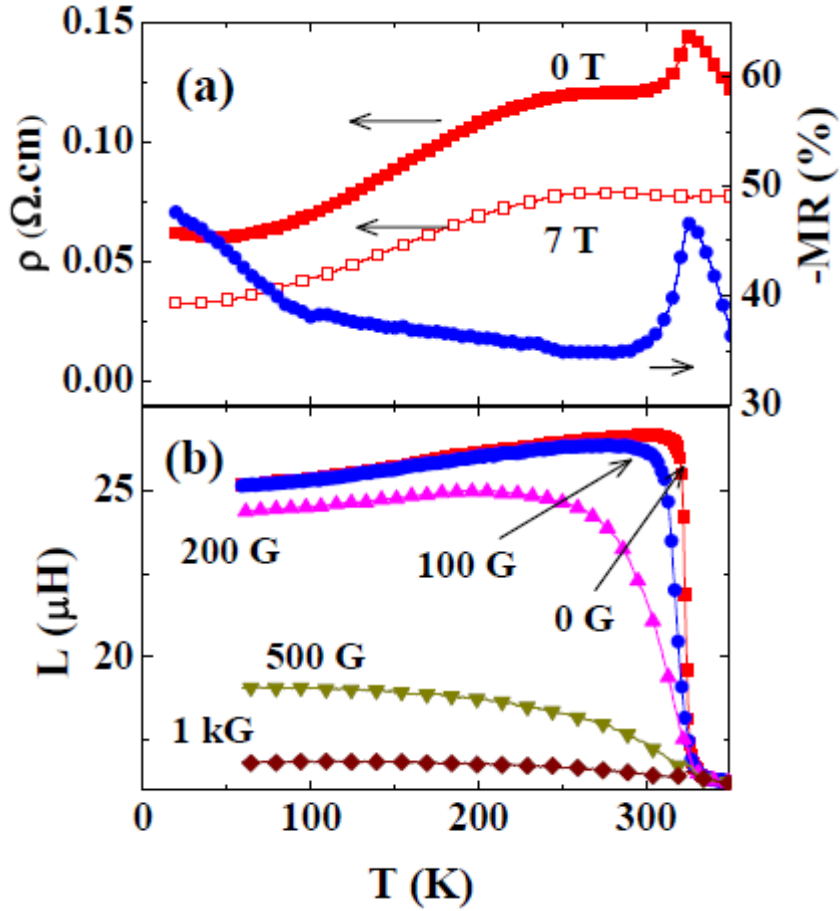

Fig. 1 (color online) (a) Temperature of the *dc* resistivity ($\rho$) of La$_{0.67}$Ba$_{0.33}$MnO$_3$ under $\mu_0H = 0$ T and 7 T (left scale) and the magnetoresistance ($\Delta\rho/\rho$) at $\mu_0H = 7$ T (right scale). (b) Temperature dependence of the *ac* inductance (L) of the sample under different *dc* bias magnetic fields measured using a solenoid.

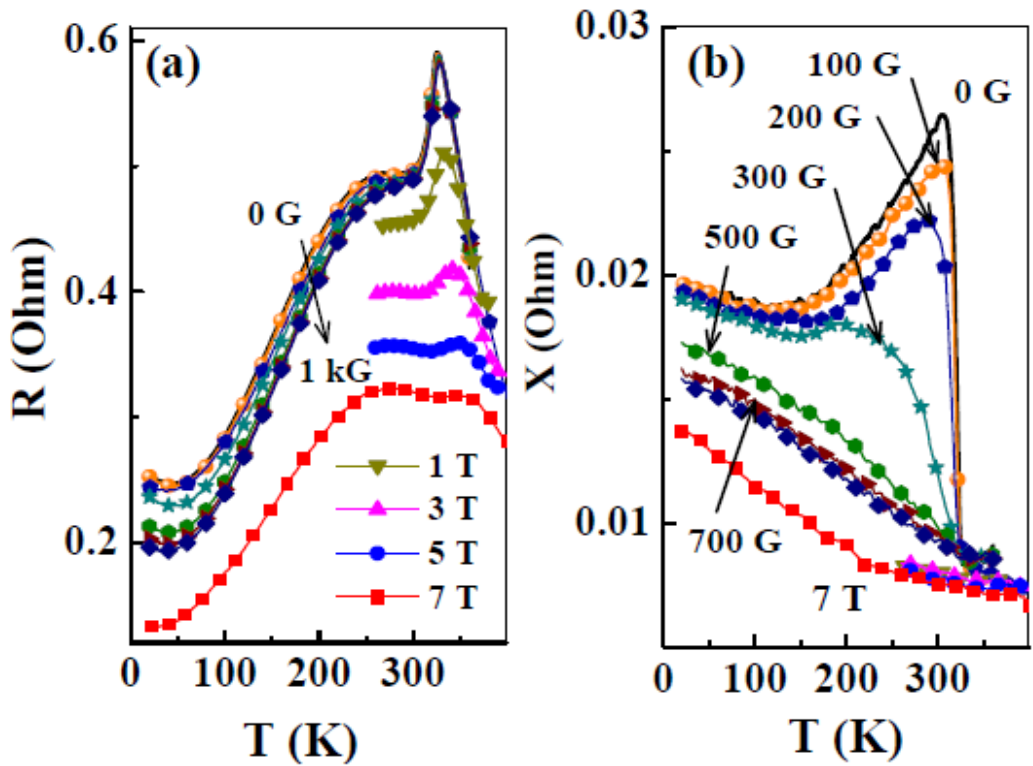

**Fig. 2** (color online) Temperature dependence of the (a) *ac* resistance (R) (b) reactance (X) under different *dc* bias fields (*H*) applied parallel to the length of the sample at *f* = 100 kHz.

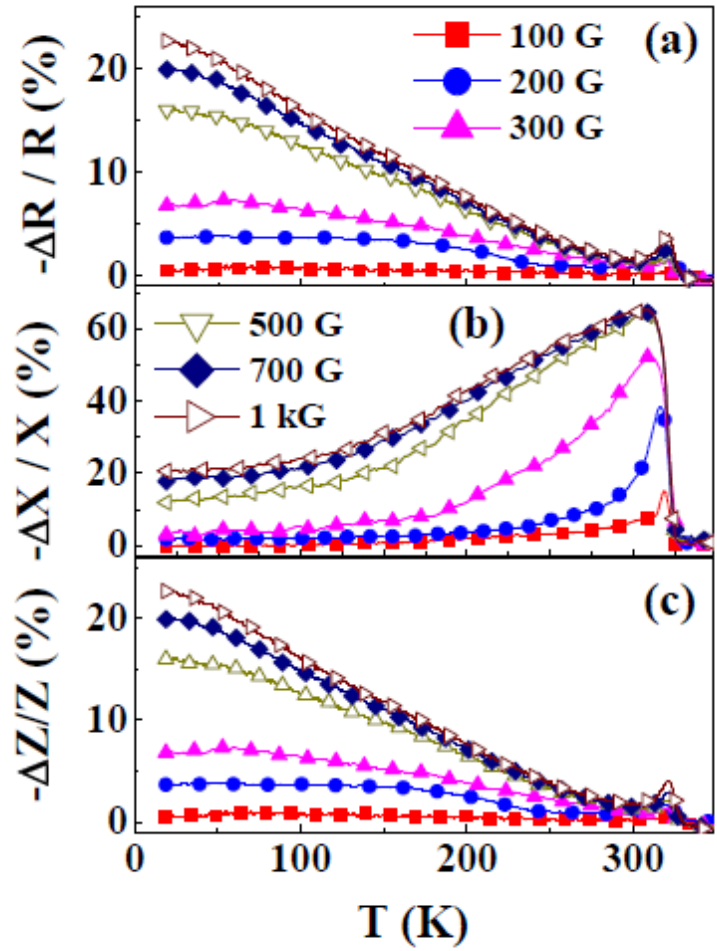

**Fig. 3 (color online) Temperature dependence of the (a) *ac* magnetoresistance (ΔR/R), (b) magnetoinductance (ΔX/X) and (c) magnetoimpedance (ΔZ/Z) under different *dc* bias magnetic fields at *f* = 100 kHz.**